\documentclass[reprint,aps,prx,amsmath,amssymb,twocolumn,superscriptaddress]{revtex4-2}
\usepackage{balance}
\usepackage{graphicx}
\usepackage[hidelinks]{hyperref}
\usepackage{amssymb}
\usepackage{slashed}
\usepackage{dcolumn}
\usepackage{amsmath}
\usepackage{bm}% bold math
\usepackage{colordvi}
\usepackage{algorithm}
\usepackage{algpseudocode}
\usepackage{multirow}
\usepackage{newtxtext,newtxmath}
\usepackage{dsfont}
\usepackage{xcolor}
\usepackage{mathbbol}

\usepackage{bm}% bold math
\usepackage{colordvi}
\usepackage{mathrsfs}
\makeatletter

\newcommand{\Rmnum}[1]{\expandafter\@slowromancap\romannumeral #1@}
\makeatother

\begin{document}   

\preprint{APS/123-QED}

\title{Spin relaxation and transport behaviors in altermagnetic systems}

\author{Y. J. Sun}

\affiliation{Department of Materials Science and Engineering and Materials Research Institute, The Pennsylvania State University, University Park, PA 16802, USA}

\author{F. Yang}
\email{fzy5099@psu.edu}

\affiliation{Department of Materials Science and Engineering and Materials Research Institute, The Pennsylvania State University, University Park, PA 16802, USA}

\author{L. Q. Chen}
\email{lqc3@psu.edu}

\affiliation{Department of Materials Science and Engineering and Materials Research Institute, The Pennsylvania State University, University Park, PA 16802, USA}

\date{\today}

\begin{abstract}
The D’yakonov-Perel’ (DP) spin-relaxation mechanism has traditionally been associated with either relativistic spin-orbit coupling, which breaks space-inversion symmetry, or inhomogeneous magnetization, which breaks both time-reversal and translational symmetries. Here, we investigate spin relaxation mechanism in altermagnetic systems which possess novel magnetic states characterized by sublattices connected through crystal-rotation symmetries and opposite spins with zero overall net magnetization and absence of spin-orbit coupling. We find that altermagnetic states exhibit DP-type spin relaxations in both strong- and weak-scattering regimes, with the spin relaxation rate decreasing to zero as the temperature approaches the critical temperature of the altermagnetic phase transition. However, the scattering time involved in this spin relaxation mechanism is not the momentum relaxation time, in contrast to the conventional DP spin relaxation. Using a kinetic approach incorporating rigorous microscopic scattering, we demonstrate that the spin Hall current is highly anisotropic and proportional to the degree of altermagnetic order.
\end{abstract}

\maketitle

\section{\label{sec:level1}Introduction}

Given the promising potential for applications in memory and computational devices, significant efforts have been devoted over the past decades to exploiting the intrinsic spins of electrons as information carriers and storage units,  an area known as spintronics~\cite{vzutic2004spintronics,vzelezny2018spin,sierra2021van,he2022topological,guo2021spintronics}, paving the way for fast information processing and high storage capacity.  One key advantage of spin-based transport and manipulation is its expected longer temporal and spatial coherence compared to charge-based transport in conventional electronics, owing to the weak coupling of spins with its environment~\cite{song2013transport, wood2022long}. The interactions of spins with environments can cause spin relaxation---a process in which an excited spin state returns to equilibrium, gradually losing its initial polarization or coherence over time~\cite{patibandla2010competing,yang2016spin,yang2015hole}. For free electrons, there are two main spin-relaxation mechanisms: Elliott-Yafet (EY)~\cite{elliott1954theory,yafet1963g,yafet1952calculation} and the D'yakonov-Perel' (DP) mechanisms~\cite{dyakonov1972spin}.  The EY mechanism arises from direct spin-flip scattering due to band mixing whereas the DP mechanism involves precessional dephasing during spin-conserving scattering, and it typically requires either spin-orbit coupling~\cite{wu2010spin}, which breaks spatial inversion symmetry, or inhomogeneous magnetization~\cite{zhang2011electron}, which disrupts both time-reversal and translational symmetries and provides an effective spin-orbit coupling SOC in $SU(2)$ gauge theory~\cite{yang2018anomalous,jia2009electrically,jia2009multiferroic,peskin2018introduction}. 

Very recently,  a distinct magnetic state known as altermagnet has been theoretically predicted and experimentally observed~\cite{vsmejkal2022emerging,bai2024altermagnetism}. This state is characterized by opposite spin sublattices connected through crystal rotation symmetries. Similar to ferromagnets, altermagnets break time-reversal symmetry; however, like antiferromagnets, they exhibit no net magnetization~\cite{vsmejkal2022emerging}. A characteristic feature of altermagnets is their spin-split electronic band structure, which alternates in momentum space. This spin splitting follows an even-parity wave character (e.g., $d$-, $g$- or $i$-wave) as the system does not break inversion symmetry~\cite{bhowal2024ferroically,reichlova2024observation,ding2024large,wei2024crystal,sodequist2024two}, in contrast to the SOC that emerges in non-centrosymmetric systems and exhibits an odd-parity-wave character (e.g., $p$- or $f$-wave).  Due to these unique properties, altermagnetic systems have been proposed as promising candidates for spintronic applications~\cite{weissenhofer2024atomistic,gonzalez2021efficient,feng2022anomalous,liao2024separation,sato2024altermagnetic,wu2010spin}. Therefore, the main objective of this work is to understand  the spin generation and relaxation mechanisms in altermagnets by studying the microscopic nonequilibrium spin dynamics.

\begin{figure}[htb]
    \centering
    \includegraphics[scale=0.6]{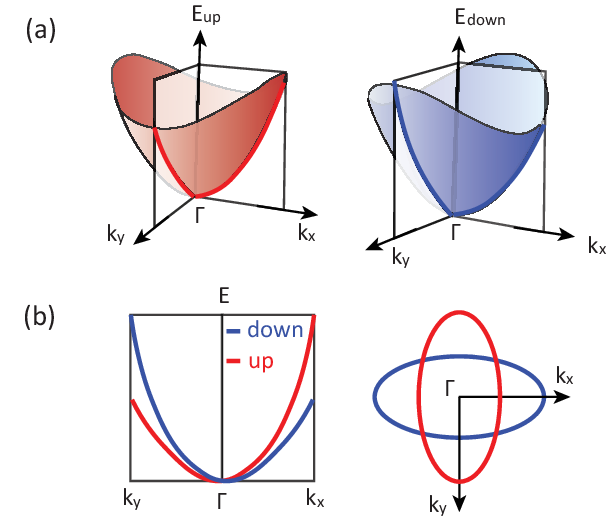}
    \caption{%\doublespacing
  Schematic illustrations of the energy bands in $d$-wave altermagnetic systems. (a) Energy dispersions for spin-up (left) and -down (right) electrons. (b) Energy dispersion along $k_x$ and $k_y$ directions (left) and the Fermi surfaces for spin-up and -down electrons (right).}
    \label{fig:intro}
\end{figure}
%%%

Here we investigate the temperature-dependent spin relaxation and the generation of spin Hall current~\cite{xiao2010berry,sinova2015spin,jungwirth2012spin,liu2022photonic,wang2021intrinsic} in altermagnetic systems using a purely microscopic kinetic model. Specifically, we consider $\beta$-MnO$_2$, a material recently predicted via first-principles calculations to exhibit altermagnetism with $d$-wave spin splitting (as illustrated in Fig.~\ref{fig:intro})~\cite{vsmejkal2022emerging,noda2016momentum}.  We derive a microscopic phase-transition theory for altermagnetic systems under a mean-field approximation and incorporate it into the kinetic spin Bloch equation for free electrons~\cite{wu2010spin}. For spin relaxation under an in-plane spin injection, we numerically and analytically analyze the relaxation process. Our results reveal that although altermagnets preserve inversion and translational symmetries, they still induce DP-type~\cite{dyakonov1972spin} spin relaxation behavior, with the spin-relaxation time following $\tau_s\propto\tau_c$ in the strong-scattering regime and $\tau_s\propto\tau_c^{-1}$ in the weak-scattering regime, where $\tau_c$ denotes a characteristic scattering time. The spin-relaxation rate decreases to zero as the temperature approaches the critical point of the altermagnetic phase transition, as expected.
For spin generation, we calculate the spin Hall current, which traditionally requires the presence of SOC~\cite{xiao2010berry,sinova2015spin}. However, in altermagnetic systems, highly anisotropic spin-dependent group velocities directly result in finite and vanishing spin Hall currents for transport along nodal and antinodal points of the $d$-wave spin splitting, respectively~\cite{vsmejkal2022anomalous}. These findings should provide insights into spin dynamics in altermagnetic systems, guiding experimental measurements and advancing spintronic applications.  Moreover, our findings demonstrate that SOC is not a prerequisite for DP-type relaxation.

\begin{figure}[bht]
    \centering
    \includegraphics[scale=0.85]{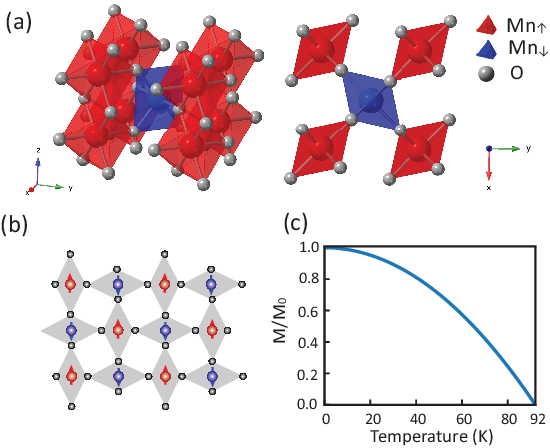}
    \caption{%\doublespacing
    (a) The unit cell of $\beta$-MnO$_2$~\cite{noda2016momentum}. (b) Alternating magnetic and atomic pattern of $d$-wave altermagnetic systems. (c) Normalized magnetization of sublattice as a function of temperature from Eq.~(\ref{eq: magnetization}) by fitting to experimentally measured $T_c=92~$K.}
    \label{fig:MnO2}
\end{figure}

\section{Models}\label{models}

Here, we present a microscopic phase-transition theory for calculating equilibrium properties and the kinetic equations for analyzing nonequilibrium spin dynamics.

{\it Phase transition and Hamiltonian.---}In altermagnetic systems, the two sublattices contain opposite magnetic moments with 
a zero net magnetization~\cite{vsmejkal2022emerging}, similar to  antiferromagnetic systems. Following the framework of antiferromagnetic phase transitions~\cite{anderson1952approximate} and using the mean-field approximation\cite{kardar2007statistical,chen2022thermodynamic}, we derive the temperature-dependent magnetization of a sublattice in a $d$-wave altermagnetic system (refer to Appendix \ref{chp:phase}): 
\begin{equation}
\label{eq: magnetization}
\langle M(T) \rangle = \langle M_{0} \rangle \Big(1- C\sum_{\bf q} \frac{n_B(\hbar\omega_{\bf q})}{\hbar\omega_{\bf q}}\Big),
\end{equation}
where $\langle M_0\rangle$ is the  magnetization at zero temperature and the second term in above equation accounts for thermal fluctuations of magnetic moments (i.e., thermal excitations of magnons). Here, $C$ is a constant and $\hbar\omega_{\bf q}$ denotes the energy spectrum of the corresponding bosonic magnon excitations in altermagnetic systems, which can be approximately as ${\bf \omega}_{\bf q}={\bf v}\cdot{\bf q}$ at long wavelength, with ${\bf v}$ being an anisotropic group velocity~\cite{PhysRevLett.133.156702,weissenhofer2024atomistic};  $n_B(x)=1/\{\exp[x/(k_BT)]-1\}$ stands for the Bose-Einstein distribution.  Recently, first-principles calculations predicted that $\beta$-MnO$_2$,  which has a tetragonal rutile
structure [Fig.~\ref{fig:MnO2}(a)], is an altermagnet. It features a rotational symmetry that connects opposite-spin sublattices, combining a two-fold spin rotation with a four-fold lattice rotation 
[Fig.~\ref{fig:MnO2}(b)], characteristic of $d$-wave altermagnetism. Experimental measurements of $\beta$-MnO$_2$ reported a transition temperature of $T_c=92~$K~\cite{zhou2018magnetic}. By fitting this transition temperature using Eq.~(\ref{eq: magnetization}),  
 we obtain the normalized sublattice magnetization as a function of temperature, as shown  in Fig.~\ref{fig:MnO2}(c). 
 
 The energy-band structure of free electrons was calculated within first-principles calculations in Ref.~\onlinecite{noda2016momentum}, showing features along $M$-$\Gamma$-$M'$ similar to those in Fig.~\ref{fig:intro}. Based on the obtained band structure along $M$-$\Gamma$-$M'$ and assuming that N'eel vector of the sublattices are aligned in the $z$-axis, an effective two-band tight-binding Hamiltonian is given by 
 \cite{vsmejkal2022giant}:
\begin{equation}
\label{eq: gen_Hamil}
H=-2t(\cos{k_x}+\cos{k_y})\sigma_0-2t_j(\cos{k_x}-\cos{k_y})\sigma_z,
\end{equation}
where $t$ represents the nearest-neighbor hopping and $t_j$ denotes an effective spin-dependent hopping by the non-relativistic alternating spin-momentum couplings and is temperature dependent, i.e., $t_j(T)/t_j(T=0)=\langle M(T)\rangle/\langle M_{0} \rangle$; $\sigma_{i=0,x,y,z}$ stand for the spin Pauli matrices.  The 
Hamiltonian above at a lightly electron-doped case can be simplified as 
\begin{equation}
\label{eq: Hamil}
H_{\textbf{k}}=\frac{\hbar^{2}k^2}{2m}\sigma_0+\gamma(T)\hbar^2k^2\cos({2\theta}_{\bf k})\sigma_z,
\end{equation}
with $m=\hbar^2/(2t)$ being effective mass and $\gamma(T)=t_j(T)/\hbar^2$ denoting the strength of alternating spin-momentum coupling.

{\it Kinetic spin bloch equations.---}
To calculate the nonequilibrium spin dynamics, we employ the kinetic spin Bloch equation (KSBE), which has been successfully applied to study spin dynamics in semiconductors~\cite{wu2010spin,weng2004hot}. In this 
microscopic approach, the response of free electrons is described by the density matrix $\rho_{\bf k}=\rho_{\bf k}^{(0)}+\delta\rho_{\bf k}(t)$ in spin space, which consists of the equilibrium part $\rho_{\bf k}^{(0)}=\left(\begin{array}{cc} f(\varepsilon_{{\bf k}\uparrow}) & 0 \\ 0& f(\varepsilon_{{\bf k}\downarrow})\end{array}\right)$ and the
nonequilibrium one $\delta\rho_{\bf k}(t)$. Here, $f(x)=1/\{\exp[(x-\mu)/(k_BT)]+1\}$ denotes the Fermi distribution, with $\mu$ being the chemical potential; $\varepsilon_{{\bf k}s=\pm}={\hbar^{2}k^2}/({2m})+s\gamma(T)\hbar^2k^2\cos{2\theta}_{\bf k}$. The KSBE for solving the nonequilibrium density matrix is written as~\cite{wu2010spin,weng2004hot} 
\begin{equation}
\label{eq: general}
{\partial_t\rho_\textbf{k}}=-i[H_{\textbf{k}}+\Sigma_{\rm HF}({\bf k}), \rho_{\textbf{k}}] -e\textbf{E}\cdot\nabla_\textbf{k} \rho_\textbf{k}
+\partial_t\rho_{\bf k}|_{\rm scat}^{\rm e-i}.
\end{equation}
where the first term on the right of the equation represents the coherent term descrbing the microscopic spin precession, with $\Sigma_{\rm HF}({\bf k})=-\sum_{\bf k'}V_{\bf kk'}\rho_{\bf k'}$ being the Hartree-Fock self-energy by Coulomb interaction $V_{\bf kk'}$~\cite{wu2010spin,weng2004hot}, the second term denotes the drive term, characterizing the response of system to
external electric field ${\bf E}$, and the last term stands for the microscopic scattering term, and for the electron-impurity scattering, it reads~\cite{wu2010spin,weng2004hot,yang2015hole}
\begin{eqnarray}
\partial_t\rho_{\bf k}|_{\rm scat}^{\rm e-i}\!=\!-n_i\pi\!\sum_{\bf k'}\!|V_{\bf kk'}|^2[(T_s\rho_{\bf k}\!-\!\rho_{\bf k'}T_s)\delta(\varepsilon_{{\bf k}s}\!-\!\varepsilon_{{\bf k}s'})\!+\!h.c.].\nonumber\\
\end{eqnarray}
Here, $V_{\bf kk'}$ is the electron-impurity interaction and $n_i$ denotes the impurity density; the projection operator
$T_{s=\pm}=(\sigma_0+s\sigma_z)/2$. It is noted that the electron-phonon and electron-electron scatterings are neglected here, 
since their effects should be minimal since the present study is focused on low temperatures ($T{\le}T_c=92~$K) and weak stimuli.   

Consequently, by solving the KSBE under different external stimuli, one can obtain the nonequilibrium dynamics of the spin density matrix and subsequently determine the temporal evolution of macroscopic properties. The expressions of the total spin polarization ${\bf S}$ and spin current ${\bf J}_{z}$ of $z$-component spin are written as
\begin{eqnarray}
{\bf S}&=&\sum_{\bf k}{\rm Tr}({\bm \sigma}\rho_{\bf k}/{2})\big/\sum_{\bf k}{\rm Tr}(\rho_{\bf k}),\label{Mspin}\\
{\bf J}^{(z)}_{\rm s}&=&\sum_{\bf k}({\bf v}_{\bf k}\sigma_z\rho_{\bf k}),\label{SHZ}
\end{eqnarray} 
with ${\bf v_k}=\partial_{\bf k}H_{\bf k}$ being the group velocity of electrons. The specific model parameters are listed in Table.~\ref{tab:table1}

\section{Results}\label{results}

In this section, we present our numerical and analytical results on spin relaxation and generation of the spin Hall current.

\begin{figure}[htb]
    \centering
    \includegraphics[scale=0.85]{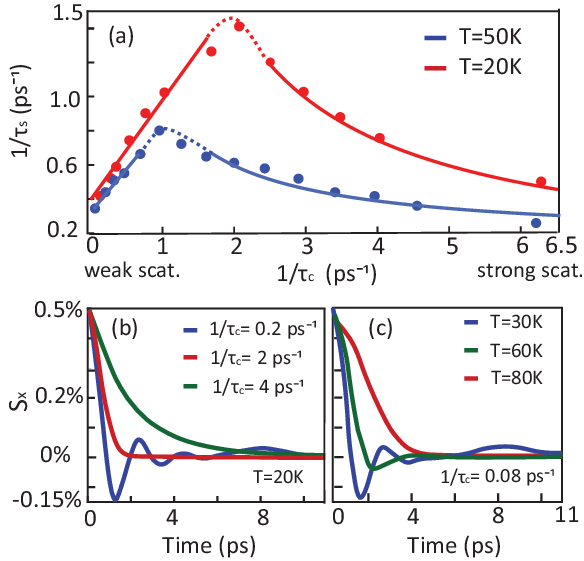}  
    \caption{%\doublespacing  
    (a) In-plane spin relaxation rate $1/\tau_s$    
    as a function of scattering rate $1/\tau_c=(n_im/\hbar^3)\int^{2\pi}_0d\theta_{\bf q}|V_{\bf q}|^2[1-\cos(2\theta_{\bf q})]$ at different temperatures. The solid curves denote the weak-scattering-regime and strong-scattering-regime relations by Eq.~(\ref{taus}). (b) The temporal evolution of the in-plane spin for strong-, mediate-
    and weak-scattering cases at $20~$K. (c) The temporal evolution of the in-plane spin for high-, mediate-
    and low-temperature cases at $1/\tau_c=0.08~$ps$^{-1}$.}
    \label{fig:relaxation}
\end{figure}

{\it Spin relaxation.---}In the field of spintronics, the spin relaxation time represents the timescale over which the excited/injected spin polarization along the chosen axis returns to its thermal equilibrium through its interaction with the environment. This parameter is critical for understanding the stability of spin states and is a key factor in determining the performance and reliability of spintronic devices~\cite{wu2010spin,sverdlov2015silicon}. To gain a deep insight into spin-relaxation mechanism in altermagnetic systems, we consider an initial state with a short-pulse in-plane (e.g., $x$-direction) spin injection from metals:
\begin{equation}
\delta\rho_{\bf k}(t=0)=\eta{R}^{\dagger}\left(\begin{array}{cc} f(\varepsilon_{{m}_{{\bf k}\uparrow}}) & 0 \\ 0& f(\varepsilon_{m{{\bf k}\downarrow}})\end{array}\right)R,
\end{equation}
with $R=\cos(\theta)\sigma_0+i\sin(\theta)\sigma_y$ and $\varepsilon_{m{\bf k}s}={\hbar^2k^2}/({2m_e})+sh_0$. We set $\theta=\pi/4$, $\eta=0.1$ and $h_0=0.1\mu$ to generate an initial $x$-direction spin polarization ${\bf S}(t=0)=(0.5\%,0,0)$. We then compute the temporal evolution of the macroscopic spin polarization without an external field. The numerical results are presented in Fig.~\ref{fig:relaxation}.

As seen from Fig.~\ref{fig:relaxation}(b), at low temperatures, the in-plane spin polarization exhibits oscillatory decay behavior in the weak scattering regime (scattering rate $1/\tau_c=0.2/$ps) and single-exponential decay in the strong scattering regime ($1/\tau_c=4/$ps). However, the fastest spin relaxation occurs at an intermediate scattering strength ($1/\tau_c=2/$ps), lying between the strong and weak scattering regimes. The characteristic spin relaxation time $\tau_s$ as a function of momentum-relaxation time is shown in Fig.~\ref{fig:relaxation}(a), and it is found that $\tau_s\propto\tau_c$ in the weak-scattering regime, and 
$\tau_s\propto1/\tau_c$  in the strong-scattering regime, exhibiting behavior similar to DP spin relaxation. Physically, this occurs because the $d$-wave spin splitting by altermagnetism induces an inhomogeneous broadening (momentum-dependent spin precessions)~\cite{wu2010spin}, which in turn triggers DP-type spin relaxation in the presence of spin-conserving scattering~\cite{sohn2024dyakonov,boross2013unified}. Thus, at weak scattering, the scattering causes the loss 
of spin coherence, resulting in $\tau_s\propto\tau_c$. In contrast,   at strong scattering, scattering inhibits the spin-precessions and consequently spin relaxation, leading to $\tau_s\propto1/\tau_c$. 

Based on this understanding, we derive an analytical solution of the spin relaxation from KSBE with $d$-wave spin-momentum coupling at both strong- and weak-scattering limits (refer to appendix \ref{chp:relaxation}):
\begin{equation}\label{taus}
S_x(t)=\begin{cases}
S_0e^{-t/(2\tau_c)}\cos\big(\sqrt{2}\gamma{k_F^2}t\big) & {\rm for~} \tau_c\gamma{k_F^2}\gg1, \\S_0e^{-2\tau_c\gamma^2k_F^4t} & {\rm for~} \tau_c\gamma{k_F^2}\ll1,
\end{cases}
\end{equation}
which shows oscillatory decay behavior in the weak scattering regime with $\tau_s\propto\tau_c$ and single-exponential decay in the strong scattering regime with $\tau_s\propto1/\tau_c$. These results are in agreement with the full numerical calculations, as shown in Fig.~\ref{fig:relaxation}(a).

It should be emphasized that the scattering time $\tau_c$ involved in the DP-type spin relaxation in altermagnetic system [$1/\tau_c=(n_im/\hbar^3)\int^{2\pi}_0d\theta_{\bf q}|V_{\bf q}|^2[1-\cos(2\theta_{\bf q})]$] is not the momentum relaxation time $\tau_p$ [where $1/\tau_p=(n_im/\hbar^3)\int^{2\pi}_0d\theta_{\bf q}|V_{\bf q}|^2(1-\cos\theta_{\bf q})$], distinguishing it from conventional DP spin relaxation, which is driven by SOC~\cite{tahan2005rashba,szolnoki2017spin}. This is because the momentum relaxation process typically concerns the odd-parity-wave distribution in momentum space. However, in altermagnetic systems, altermagnetism leads to an even-parity spin-momentum coupling, in contrast to the odd-parity spin-momentum coupling by SOC. The high-parity-order scattering effects typically do not manifest in experimental measurements; however, they become visible in the spin relaxation behavior in altermagnetic systems because of their unique symmetry-breaking properties.

Moreover,  for a fixed scattering time, the spin relaxation decreases with increasing temperature and eventually approaches zero as the temperature nears the critical point of the altermagnetic phase transition. Notably, the crossover point [$\tau_c\gamma(T)k_F^2\sim1$] 
 between strong- and weak-scattering regimes shifts toward lower $1/\tau_p$ at higher temperatures. Therefore, 
as the temperature increases, one can expect to observe a crossover in spin relaxation from weak- to strong-scattering regime behavior, as shown in Fig.~\ref{fig:relaxation}(c), offering a potential detection scheme for experimental validation.
 
\begin{figure}[htb]
    \centering
    \includegraphics[scale=0.44]{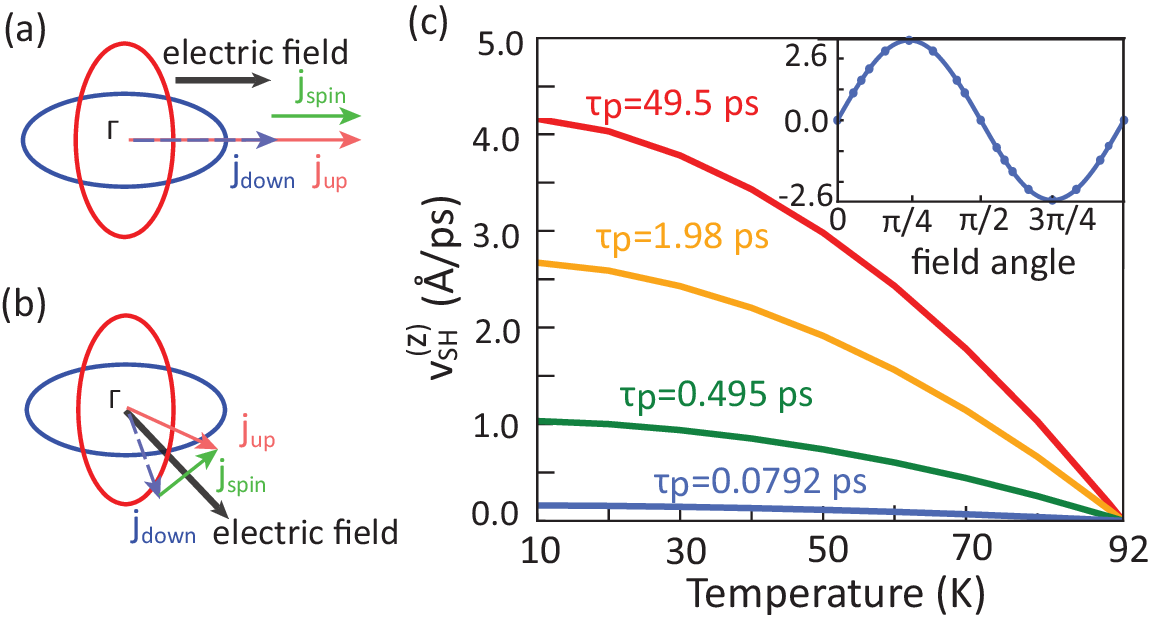}
    \caption{%\doublespacing
    Schematic illustrations of generation of spin current and spin Hall current 
    for an electric field applied along (a) anti-nodal and (b) nodal points of the $d$-wave spin splitting. (c) Spin-Hall drift velocity ${v}^{(z)}_{\rm SH}$
    as a function of temperature at different momentum-relaxation times $\tau_p$. The inset shows the ${v}^{(z)}_{\rm SH}$ versus the electric-field angle.}
    \label{fig:spinHall}
\end{figure}

{\it Spin Hall current.---}Apart from controlling spin relaxation, generating a stable spin current that carries information is another key process for spintronic applications. In potential memory devices, a spin current perpendicular to the charge current (spin Hall current) is preferred due to its stable response in reading and writing processes \cite{brataas2012current, ralph2008spin}. Traditionally, the generation of spin Hall current requires the presence of SOC. However, we show that in altermagnetic systems without SOC, highly anisotropic spin-dependent group velocities can give rise to the spin Hall current~\cite{PhysRevLett.130.216701}.

Specifically, in the presence of a $d$-wave spin splitting, the group velocities of spin-up and -down electrons read
\begin{equation}\label{vks}
{\bf v}_{{\bf k}s}=\hbar{\bf k}/m+s\gamma\hbar{\bf k}\cos(2\theta_{\bf k})+s2\gamma\hbar{k}\sin(2\theta_{\bf k}){\bf e}_{\theta_{\bf k}},
\end{equation}
which exhibit strong anisotropy and include a component (the last term) perpendicular to the momentum ${\bf k}$. As shown in Fig.~\ref{fig:spinHall}(a), for electric field applied along the anti-nodal points ($\theta_{\bf k}=n\pi/2$ where $n$ is an integer) of the $d$-wave spin splitting,  ${\bf v}_{{\bf k}\uparrow}$ and ${\bf v}_{{\bf k}\downarrow}$ despite having different magnitudes, remain aligned in the same direction as the momentum ${\bf k}$, resulting in a spin current parallel to the charge current. Nevertheless, for electric field applied along the anti-nodal points ($\theta_{\bf k}=\pi/4+n\pi/2$) of the $d$-wave spin splitting, as shown in  Fig.~\ref{fig:spinHall}(b), the spin-up and spin-down electron flows deviate from the total charge current direction in opposite directions, leading to the generation a spin-Hall current.  Based on this understanding,  we present a simple derivation of the transverse (perpendicular to ${\bf E}$) component in ${\bf J}_s^{(z)}$ under relaxation-time approximation:
\begin{equation}\label{SHC}
{J}^{(z)}_{\rm SH}=n{v_{\rm SH}^{(z)}}\approx-{8eEn\tau_p}\gamma \sin{2\theta},
\end{equation}
with $n$ being the charge density and $v_{\rm SH}^{(z)}$ being the drift velocity of spin Hall current; $\theta$ denoting the field angle.

For the full numerical calculation, we apply a finite external electric field and solve for the steady-state in the presence of rigorous microscopic scattering, in contrast to the relaxation-time approximation used in existing literature~\cite{sato2024altermagnetic}. 
 The numerical results of the spin-Hall current of the $z$-component spins are plotted in Fig.~\ref{fig:spinHall}(c). As shown in the inset, the induced spin Hall current reaches its maximum when the electric field is applied along the antinodal points and vanishes when applied along the nodal points, consistent with the analysis above. Moreover, the spin Hall current decreases with increasing temperature and eventually approaches zero as the temperature nears the critical point of the altermagnetic phase transition, indicating its origin from altermagnetism. Additionally, the spin Hall current is suppressed with increasing momentum relaxation scattering, as expected. All these numerical results incorporating rigorous microscopic scattering are in agreement with the simple derivation in Eq.~(\ref{SHC}) under relaxation-time approximation. 

Furthermore, it is important to emphasize that in this scenario, the breaking of spin-up and spin-down density degeneracy can lead to the generation of a charge Hall current (refer to Appendix~\ref{chp:spinHall}). This offers a potential detection scheme for experimental validation through transport and/or optical measurements.

\section{Summary}

In summary, using a purely microscopic kinetic model with rigorous microscopic scattering,  we numerically and analytically investigated the spin relaxation and generation of spin Hall current in a $d$-wave altermagnetic system, $\beta$-MnO$_2$.  Our results reveal that although altermagnets preserve inversion and translational symmetries, they are still able to induce DP-type spin relaxation and/or spin-Hall current,   demonstrating that SOC or inhomogeneous magnetization is not the prerequisite for DP-type relaxation and/or spin-Hall current.

For experimental validation, we demonstrate that as the temperature increases, spin relaxation undergoes a crossover from the weak- to strong-scattering regime, exhibiting distinct temporal evolution behaviors. It should be emphasized that the spin relaxation cannot occur in ferromagnetic or antiferromagnetic systems without the presence of SOC. These serve as a means to validate potential altermagnetic candidates by analyzing the spin relaxation behavior of a system.
On the other hand, we show that the spin-Hall current reaches its maximum when the electric field is applied
along the antinodal points and vanishes when applied along the
nodal points. Breaking of spin-up and spin-down density degeneracy in this case can lead to the generation of a charge Hall current, offering a potential detection scheme through transport and/or optical measurements~\cite{farajollahpour2024light}. 

Moreover, by symmetry analysis~\cite{yang2017gapped}, combining a conventional spin-singlet $s$-wave superconductor with the $d$-wave altermagnetism can generate a spin-triplet odd-frequency $d$-wave pairing at the interface~\cite{fukaya2025josephson}, which then can diffuse into both materials. The revealed spin relaxation and spin-Hall current generation here through high-order-parity scattering is expected to significantly influence on this diffusion, e.g., long-range spin–triplet correlations~\cite{bergeret2001long} and spin supercurrent~\cite{linder2015superconducting}.

This work highlights the potential of altermagnetic materials like $\beta$-MnO$_2$ for advancing spintronics, leveraging their unique symmetry-breaking properties to achieve SOC-independent spin relaxation and spin Hall effects. These findings offer new perspective to explore novel spintronic devices where temperature, impurity scattering, and material design converge to control spin-based functionalities.

\begin{acknowledgments}

Y.J.S. was supported by the National Science Foundation  under Grants No. DMR-2133373.
 F.Y. and L.Q.C. acknowledge support from the US Department of Energy, Office of Science, Basic Energy Sciences, under Award
Number DE-SC0020145 as part of Computational Materials
Sciences Program. F.Y. and L.Q.C. also appreciate the generous support from the Donald W. Hamer Foundation through a Hamer Professorship at Penn State.
 
\end{acknowledgments}

\appendix
%\begin{appendix}
\section{Phase-transition model}
\label{chp:phase}

In our model, the temperature dependence of the alternating spin-momentum coupling $\gamma(T)$ in free-electron Hamiltonian [Eq.~(\ref{eq: Hamil})] follows the relation $\gamma(T)/\gamma(0) = \langle M(T)\rangle /\langle M_0 \rangle$. To quantitatively describe this dependence, we formulate the phase transitions by considering the magnon thermal excitation in altermagnetic systems, following previous works. In particular, magnons, as bosonic quasiparticles, are regarded as spin waves arising from fluctuating magnetic moments. As temperature increases, fully ordered magnetic moments become frustrated due to the thermal excitation of magnons. To capture this behavior, we employ the Heisenberg model for a system with two sets of spin sub-lattices, which is written as~\cite{auerbach2012interacting} 
\begin{eqnarray}
    H&=&J\sum_{i,\delta}\boldsymbol{S}_{A,i}\cdot\boldsymbol{S}_{B,i+\delta}+J\sum_{j,\delta}\boldsymbol{S}_{B,j}\cdot\boldsymbol{S}_{A,j+\delta}\nonumber\\
    &=&J\sum_{i,\delta}\Big({S}_{A,i}^{+}\!\cdot\!{S}_{B,i+\delta}^{-}\!+\!{S}_{A,i}^{-}\!\cdot\!{S}_{B,i+\delta}^{+}\!+\!2{S}_{A,i}^z\!\cdot\!{S}_{B,i+\delta}^z\Big).~~~~~
\label{eq:Sladder}
\end{eqnarray}
where $A$, $B$ label two sets of the sub-lattice; $i$, $j$ are the lattice indices, and $\delta$ denotes the nearest neighbor; $\boldsymbol{S}$ represents the spin orperator; $J$ is the exchange  interaction strength, which is positive for an antiferromagnetic system.

For an altermagnetic system with opposite neighboring magnetic moments, assuming the sub-lattice magnetization is along the $z$ axis, we employ the standard Holstein-Primakoff transformations~\cite{PhysRev.58.1098}:
$S_{A,i}^z=S-a_i^\dag a_i$ and $S_{B,i}^z=-S+b_i^\dag b_i$;  $a_i={S}_{A,i}^+/\sqrt{2S}$ and $a_i^{\dag}={S}_{A,i}^-/\sqrt{2S}$; $b_i={S}_{B,i}^-/\sqrt{2S}$, and $b_i^{\dag}={S}_{B,i}^+/\sqrt{2S}$. Here,  $a^\dag$ and $a$ ($b^\dag$ and $b$) are creation and annihilation spin operators for $A$ ($B$) sub-lattice. Then, the Hamiltonian in Eq.~(\ref{eq:Sladder}) becomes
\begin{eqnarray}
H&=&2JS\sum_{i}\left(a_{i}^{\dag}a_{i}+b_{i}^{\dag}b_{i}\right)+
2JS\sum_{i,\delta}\left(a_{i}b_{i+\delta}+
a_i^\dag b_{i+\delta}^{\dag} \right) \nonumber \\
&=&2JNS\sum_{\boldsymbol{q}}\Big[\left(a_{\boldsymbol{q}}^{\dag}a_{\boldsymbol{q}}+b_{\boldsymbol{q}}^{\dag}b_{\boldsymbol{q}}\right)
\!+\!\gamma_{\boldsymbol{q}}\Big( a_{\boldsymbol{q}}b_{\boldsymbol{q}}+a_{\boldsymbol{q}}^{\dag}b_{\boldsymbol{q}}^\dag
\Big)\Big],\label{eq:Sboson}
\end{eqnarray}
where $\gamma_{\boldsymbol{q}} = \sum_{\delta}e^{i\boldsymbol{q}\cdot \boldsymbol{\delta}}$ is the parameter related to the Fourier transformation to momentum space.

After diagonalizing the Hamiltonian above by applying the Bogoliubov transformation~\cite{transform}, one obtains 
\begin{equation}
    H=\sum_{\boldsymbol{q}} \omega_{\boldsymbol{q}} \left( \alpha_{\boldsymbol{q}}^\dag \alpha_{\boldsymbol{q}} 
    +\beta_{\boldsymbol{q}}^\dag \beta_{\boldsymbol{q}} \right),
\end{equation}
where $\omega_{\boldsymbol{q}}=2JSN\sqrt{1-\gamma_{\boldsymbol{q}}^2}$ is the excitation energy of magnons; $\alpha_{\bm q}$ and $\beta_{\bm q}$ denote the annihilation operator of magnons. Particularly, by expanding $\omega_{\boldsymbol{q}}=2JSN\sqrt{1-\gamma_{\boldsymbol{q}}^2}$ in the long-wavelength limit, one finds a gapless energy spectrum $\omega_{\boldsymbol{q}}= \boldsymbol{v}\cdot\boldsymbol{q}$ as a consequence of the Nambu-Goldstone theorem associated with the spontaneous breaking of continuous spin-rotational symmetry~\cite{auerbach2012interacting}.
 In an altermagnetic system, the  magnitude $|\boldsymbol{v}|$ of velocity varies in different directions~\cite{PhysRevLett.133.156702}.

% see u and v values \href{https://physics.stackexchange.com/questions/268268/bogoliubov-transformation-is-not-unitary-transformation-correct}{here}

In specific calculations, the magnetization in a sub-lattice can be derived as\cite{bloch1963aspects,PhysRevB.37.9761,mubayi1969phasemubayi1969phase}:
\begin{eqnarray}
\langle M(T) \rangle & =&\sum_i\langle{S_{A,i}^Z}\rangle=NS -N\sum_{\boldsymbol{q}}\langle a_{\boldsymbol{q}}^\dag a_{\boldsymbol{q}} \rangle  \nonumber \\
&=&NS-N\sum_{\boldsymbol{q}} \frac{n_B(\hbar\omega_{\boldsymbol{q}})}{\sqrt{1-\gamma_{\boldsymbol{q}}^2}}  \nonumber \\
&=&\langle M_0\rangle \Big(1- C \sum_{\boldsymbol{q}} \frac{n_B(\hbar\omega_{\boldsymbol{q}})}{\hbar\omega_{\boldsymbol{q}}}\Big)  \nonumber \\
&=&\langle M_0\rangle \Big(1-\tilde{C}\int_0^{\omega_c}{\frac{\omega_{\boldsymbol{q}}e^{-\omega_{\boldsymbol{q}}/(k_BT)}}{1-e^{-\omega_{\boldsymbol{q}}/(k_BT)}}}d\omega_{\boldsymbol{q}}\Big)  \nonumber \\
&=&\langle M_0\rangle \Big\{ 1-\tilde{C}\sum_{n=1}^{\infty}\frac{k_BT}{n^2}\Big[-e^{-\frac{\omega_{\boldsymbol{q}}}{k_BT}}\Big(\frac{\omega_{\boldsymbol{q}}n}{k_BT}+1\Big)\Big]_{0}^{\omega_{c}}\Big\}  \nonumber \\
&=&\langle M_0\rangle\Big\{1\!-\!\sum_{n}\frac{T^2}{n^2T_1^2}\Big[-e^{-\frac{nT_2}{T}}\Big(\frac{nT_2}{T}\!+\!1\Big)\!+\!1\Big]\Big\},\label{MT}
\end{eqnarray}
in which only two undetermined parameters $T_1$ and $T_2$ as fitting parameters. Here, $\omega_c$ denotes the cutoff energy; the prefactors $C$ and $\tilde{C}$ are normalized coefficients. This phase transition model is a rough approximation, but it does not impact the main conclusions regarding free electrons in this study.

\section{Analytical solution for spin relaxation}
\label{chp:relaxation}

In this part, we present the analytical solution for spin relaxation. For the free decay in a nonequilibrium initial state, the KSBE without the driving term is written as:
\begin{equation}\label{saa}
    \partial_t \rho_{\boldsymbol{k}}+i[\Omega_{\boldsymbol{k}} \sigma_z, \rho_{\boldsymbol{k}}]=\partial_t\rho_{\bf k}|_{\rm scat}^{\rm e-i},
\end{equation}
where $\Omega_{\boldsymbol{k}}=\gamma{k}^2 \cos{(2\theta_{\boldsymbol{k}})}$, and we set $\hbar=1$ for convenience.  We focus on the states near Fermi level $\boldsymbol{k}\sim \boldsymbol{k}_F$ because of their large contribution to the dynamic behaviors. Considering the $d$-wave symmetry character of the system, one can expand $\rho_{\boldsymbol{k}}=\sum_{l} \rho_{{k}}^l \cos{(l\theta_{\boldsymbol{k}})}$, where $l\in \{0, \pm2\}$ while we have neglected higher order terms. Then, Eq.~(\ref{saa}) becomes two simplified equations:
\begin{eqnarray}
        \label{eq:rho1}
        &&\partial_t \rho_{k_F}^{l=0}+i[\gamma {k_F}^2 \sigma_z, \rho_{k_F}^{l=\pm2}]=-\frac{\rho_{k_F}^{l=0}}{\tau_{{k_F}}^{l=0}},\\ 
        \label{eq:rho2}
        &&\partial_t \rho_{k_F}^{l=\pm2}+\frac{i}{2}[\gamma {k_F}^2 \sigma_z, \rho_{{k_F}}^{l=0}]=-\frac{\rho_{k_F}^{l=\pm2}}{\tau_{k_F}^{l=\pm2}},
\end{eqnarray}
where $\frac{1}{\tau_{k_F}^{l}}=n_im\int{d}\theta_{\bf k_F'}|V_{\bf k_F-k_F'}|^2\{1-\cos[l(\theta_{\bf k_F}-\theta_{\bf k_F}')]\}$~\cite{wu2010spin}. Consequently, Eq.~(\ref{eq:rho1}) can be written as
\begin{equation}
\Big(\partial_t+\frac{1}{\tau_{{k_F}}^{l=2}}\Big)\rho_{k_F}^{l=\pm2}+\frac{i}{2}[\gamma {k_F}^2 \sigma_z, \rho_{k_F}^{l=0}]=0.
\end{equation}
Multiplying $(\partial_t+1/\tau_{{k_F}}^{l=2})$ to  both sides of Eq.~(\ref{eq:rho1}), by expanding $\rho_{k_F}^l=\sum_{i=x,y,z,0}\rho_{{k_F},i}^l\sigma_i$ and only considering the spin behaviors on $x$ direction (i.e., the component $\rho^l_{{k_F},x}$), one gets
\begin{eqnarray}
        &&\Big(\partial_t+\frac{1}{\tau_{k_F}^{l=2}}\Big)\partial_t \rho_{k_F,x}^{l=0}\sigma_x+\frac{\gamma^2k^4_F}{2}\left[\sigma_z, [\sigma_z, \rho_{k_F,x}^{l=0}\sigma_x]\right]=0 \nonumber\\
        &&\Rightarrow\Big(\partial_t^2+\frac{\partial_t}{\tau_{k_F}^{l=2}}+2\gamma^2k^4_F\Big)\rho_{{k_F},x}^{l=0}=0.\label{SFF}
\end{eqnarray}
Consequently, with the equation above, the equation of motion macroscopic spin behaviors [i.e., $S_x(t)$] from Eq.~(\ref{Mspin}) reads
\begin{equation}
\Big(\partial_t^2+\frac{\partial_t}{\tau_{k_F}^{l=2}}+2\gamma^2k^4_F\Big)S_x(t)=0.
\end{equation}
The characteristic roots of this equation are $-1/(2\tau_{k_F}^{l=2})\pm\sqrt{{1/(2\tau_{k_F}^{l=2}})^2-2\gamma^2k^4_F}$. When the scattering is weak $1/\tau^{l=2}_{k_F}\rightarrow 0$, the in-plane spins relax in an oscillatory decaying manner with $e^{-t/(2\tau_{k_F}^{l=2})}\cos{\big(\sqrt{2\gamma^2k^4_F}t\big)}$, where the spin relaxaion time $\tau_s$ is proportional to the scattering relaxation time $\tau_{{k_F}}^{l=2}$. When the scattering is strong, i.e., the scattering rate $1/\tau_{k_F}^{l=2} \rightarrow \infty$, the oscillation behavior dies down and the long-time relaxation behavior follows the single exponential decay behavior as $e^{-2\gamma^2{{k_F}}^4\tau_{k_F}^{l=2}}$, leading to an inverse relation between $\tau_s$ and $\tau_{{k_F}}^{l=2}$. Clearly, the scattering time $\tau_{k_F}^{l=2}$ involved in the spin relaxation here is not the momentum relaxation time $\tau_{k_F}^{l=1}$, distinguishing it from conventional DP spin relaxation~\cite{tahan2005rashba,szolnoki2017spin}. 

\section{Spin Hall current}
\label{chp:spinHall}
In this section, we present analytical solution for the spin-Hall current in an altermagnetic system.  Assuming the density matrix $\rho_{\bf k}=\left(\begin{array}{cc} f_{\bm k\uparrow}^0+\delta f_{\bm k \uparrow} & 0 \\ 0& f_{\bm k\downarrow}^0+\delta f_{\bm k \downarrow}\end{array}\right)$ where $f_{{\bf k}s}^0$ and $\delta{f_{{\bf k}s}}$ represent the equilibrium and nonequilibrium parts of electron distribution for spin $s$, the KSBE under the relaxation time approximation can be simplified as:
\begin{equation}
    \frac{\partial \left(\delta f_{\boldsymbol{k}s}\right)}{\partial t}+e{\bf E} \cdot {\bm \nabla}_{{\bf k}} f_{\boldsymbol{k}s}^0 = -\frac{ \delta f_{{\bf k}s}}{\tau_{{p}}},
\end{equation}
which is similar to the spin-dependent Boltzmann equation. Here,  $\tau_{p}$ denotes the momentum relaxation time ($\tau_{k_F}^{l=1}$). At the steady state, one can get the solution of nonequilibrium part: $\delta f_{\boldsymbol{k}s}=\tau_{p}\left(e\boldsymbol{E}\cdot \boldsymbol{v_{\boldsymbol{k}s}}\right)\partial_{\varepsilon_{{\boldsymbol{k}}s}}f_{\boldsymbol{k}s}^0$. Consequently, by substituting the velocity of spin-up and spin-down electrons with $\boldsymbol{v}_{\boldsymbol{k}, \uparrow(\downarrow)}=\hbar |k|[(1/m^*\pm 2\gamma)\cos{\theta}_{\bf k}, (1/m^*\mp2\gamma)\sin{\theta}_{\bf k}]$, from Eq.~(\ref{SHZ}), focusing mainly on electrons near the Fermi level, the spin-Hall current is given by 
\begin{widetext}
\begin{eqnarray}
        \boldsymbol{J}^{(z)}_{\text{SH}}&=&{\bf J}_{s}^{(z)}\cdot{(-\sin\theta{\bf e_x}+\cos\theta{\bf e_y})}=\Big(\sum_{\boldsymbol{k}} {\delta}f_{\boldsymbol{k},\uparrow} \boldsymbol{v}_{\boldsymbol{k},\uparrow}-\sum_{\boldsymbol{k}} {\delta}f_{\boldsymbol{k},\downarrow} \boldsymbol{v}_{\boldsymbol{k},\downarrow}\Big)(-\sin\theta{\bf e_x}+\cos\theta{\bf e_y})\nonumber\\
&=&|\boldsymbol{E}|e\tau_{\boldsymbol{k}}\hbar^2\boldsymbol{k}_F^2\frac{2m}{\hbar^2}\int_{0}^{2\pi} \sum_{s=\pm1}\frac{sd\theta_{\boldsymbol{k}}}{1+2ms\gamma\cos{2\theta_{\boldsymbol{k}}}}
\left[(1/m+s2\gamma)\cos{\theta_{\boldsymbol{k}}}\cos{\theta}+(1/m-2s\gamma)\sin{\theta_{\boldsymbol{k}}}\sin{\theta}\right] \nonumber\\
&&\times\left[-(1/m+s2\gamma)\cos{\theta_{\boldsymbol{k}}}\sin{\theta}+(1/m-2s\gamma)\sin{\theta_{\boldsymbol{k}}}\cos{\theta}\right] \nonumber \\
&=&-|\boldsymbol{E}|e\tau_{\boldsymbol{k}}\hbar^2\boldsymbol{k}_F^2\frac{2m}{\hbar^2}\int_{0}^{2\pi} \sum_{s=\pm1}\frac{sd\theta_{\boldsymbol{k}}/2}{1+2ms\gamma\cos{2\theta_{\boldsymbol{k}}}}[(1/m^2+4\gamma^2)\cos{2\theta_{\boldsymbol{k}}}+4s\gamma/m]\sin{2\theta} \nonumber
\\
&=&-|\boldsymbol{E}|e\tau_{\boldsymbol{k}}\hbar^2\boldsymbol{k}_F^2\frac{2m}{\hbar^2}\int_{0}^{2\pi} \sum_{s=\pm1}\frac{d\theta_{\boldsymbol{k}}/2}{1+2ms\gamma\cos{2\theta_{\boldsymbol{k}}}}[\frac{(1/m^2+4\gamma^2)}{2m\gamma}(2ms\gamma\cos{2\theta_{\boldsymbol{k}}}+1)-\frac{(1/m^2+4\gamma^2)}{2m\gamma}+\frac{4\gamma}{m}]\sin{2\theta}
\nonumber\\
&=&-|\boldsymbol{E}|e\tau_{\boldsymbol{k}}\hbar^2\boldsymbol{k}_F^2\frac{m}{\hbar^2}\sin{2\theta}\sum_{s=\pm1}[2\pi\frac{(1/m^2+4\gamma^2)}{2m\gamma}+\Big(-\frac{(1/m^2+4\gamma^2)}{2m\gamma}+\frac{4\gamma}{m}\Big)\frac{2\pi}{\sqrt{1-4m^2\gamma^2}}],\label{SHFF}
\end{eqnarray}
\end{widetext}
which reduces to Eq.~(\ref{SHC}) at weak alternating
spin-momentum coupling , i.e., $m\gamma\ll1$. The equation above, derived under the relaxation-time approximation, suggests that in $d$-wave altermagnetic systems, the spin Hall current exhibits a $d$-wave anisotropy. This result is consistent with our full numerical calculations, which incorporate rigorous microscopic scattering. As expected, the magnitude of the spin Hall current is proportional to the altermagnetic parameter $\gamma(T)$ and is therefore temperature-dependent.

\section{Charge Hall current}
\begin{figure}[thb]
    \centering
    \includegraphics[scale=0.39]{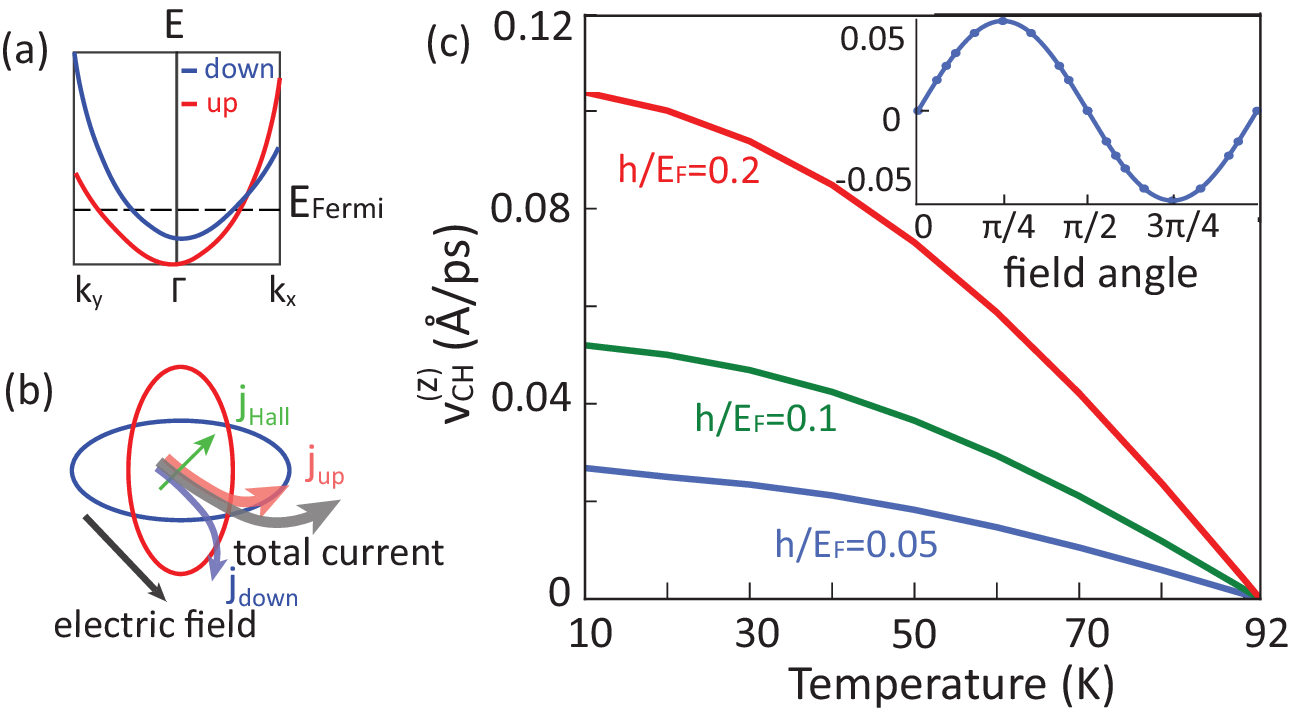}
    \caption{%\doublespacing
    (a) Energy dispersion of a $d$-wave altermagnetic system with the
breaking of spin-up and -down degeneracy by Zeeman
field.
    (b)    
    Schematic illustration of the generation of charge Hall  current.  (c) Temperature dependence of  charge Hall electron velocity at different Zeeman fields. The inset shows the dependence on the electric field angle.}
    \label{fig:chargeHall}
\end{figure}

It is known that 
in the presence of spin Hall current,  the breaking of spin-up and -down density degeneracy can lead to the generation of a charge Hall current. In this part, we consider a breaking of free-electron spin-up and -down degeneracy by an effective weak Zeeman field ($s$-wave), which can be induced through the proximity effect to ferromagnetism,  disorder-induced small net magnetization or external magnetic field.  In this situation,  an energy splitting is introduced between the spin-up and spin-down states, as illustrated in Fig.~\ref{fig:chargeHall}(a). The initial electron distribution at this case reads $\rho_{\textbf{k}s}^0=1/[e^{(\varepsilon_{{\bf k}s}+h\sigma_z-\mu)/(k_BT)}+1]$,
where $h$ is the introduced Zeeman energy.  As depicted on Fig.~\ref{fig:chargeHall} (b), the energy splitting between spin-up and spin-down states leads to unequal magnitudes of spin-up and spin-down currents. This imbalance results in the generation of a charge Hall current when an electric field is applied  along the nodal points [inset of Fig.~\ref{fig:chargeHall}(c)]. Analogous to the spin Hall effect, this charge Hall effect disappears as the system undergoes a phase transition at the critical temperature [Fig.~\ref{fig:chargeHall}(c)] as expected, and the larger energy split leads to higher charge Hall current with similar temperature dependence and electric field angle dependence. 
\balance 
%\end{appendix}
%\newpage
\begin{table}[h]%The best place to locate the table environment is directly after its first reference in text

\caption{\label{tab:table1}
Specific model parameters used in the simulations. We consider a lightly doped case with fermi energy $\text{E}_F$ slightly above the bottom of the conduction band. $m^*$ and $\gamma_0$ are fitted to the DFT results of conduction-band energy spectrucm~\cite{noda2016momentum} at 0~K. Here we consider a short-range electron-impurity interaction in the numerical calculations, and the impurity density $n_i$ and the short-range electron-impurity interaction strength $V_i$ are chosen with respect to a wide range of momentum relaxation time between 0.05 ps to 50 ps. $T_1$ in Eq.~(\ref{MT}) is fitted to experimental phase-transition temperature (92~K) in Ref.~\cite{zhou2018magnetic} with assuming an empirical $T_2$. $m_e$ denotes the free electron mass.
}
\begin{ruledtabular}
\begin{tabular}{lllll}
\textrm{KSBE model}&
\textrm{ }&
\textrm{Phase transition model}&
\textrm{ }&
\textrm{ }\\
\colrule
 $\text{E}_F$ (eV)& 0.02 & $T_1$ (K) & 117.98 & \\
%$E_{\text{cut-off}}$~(eV) & 0.2 \\
$m^*$ (eV$\cdot$ps$^2$/$\mathring{\text{A}}$)& $0.8445m_e$ & $T_2$ (K)& 1000&\\
$\gamma_{0}$~($\mathring{\text{A}}$/eV$\cdot$ps$^2$) & $0.0306/m_e$ & & &\\
$n_i|V_i|^2$~(eV$^2\mathring{\text{A}}$$^{3}$) & [0.006, 0.18] & & & \\
\end{tabular}
\end{ruledtabular}
\end{table}

\section{Computation details}
All numerical calculations  are based on the equations in Sec.~\ref{models}, and are solved by Runge-Kutta-Fehlberg (RK45) method using package SciPy’s \textit{solve\_ivp} function in Python. The Fermi energy ${E}_F$ is set to be slightly above the conduction-band bottom, and the  high-energy cutoff in our numerical simulation $E_{\rm cut}$ is $10E_{F}$.   To ensure convergence in response to an applied electric field, we simplify the equation to its quadratic term:
\begin{equation}
\label{eq: E}
e\textbf{E}\cdot\nabla_\textbf{k} \rho_\textbf{k}=e\textbf{E}\cdot\nabla_{\textbf{k}}\rho_{\textbf{k}}^0+\tau_{p}(e\textbf{E}\cdot\nabla_\textbf{k})^2 \rho_{\textbf{k}}^0,
\end{equation}
where $\rho_\textbf{k}^0$ is the initial-state density matrix of electrons following the equilibrium Fermi distribution and $\tau_p$ is the momentum-relaxation time for quadratic electric field response. Other used model parameters are listed in Table.~\ref{tab:table1}

\providecommand{\noopsort}[1]{}\providecommand{\singleletter}[1]{#1}%

\end{document}